\documentclass[11pt,a4paper]{article}
\usepackage[hyperref]{acl2020}
\usepackage{times}
\usepackage{latexsym}
\usepackage{graphicx}

\usepackage{microtype}

\aclfinalcopy 
\def\aclpaperid{2} 

\title{Representation Learning for Discovering Phonemic Tone Contours}

\author{Bai Li$^{1,2}$, Jing Yi Xie$^{1}$, Frank Rudzicz$^{1,2,3}$ \\
    $^1$ University of Toronto, Toronto, Canada \\
    $^2$ Vector Institute, Toronto, Canada \\
    $^3$ St Michael's Hospital, Toronto, Canada \\
  {\tt \{bai, frank\}@cs.toronto.edu, jingyi.xie@mail.utoronto.ca} 
  } 

\date{}

\begin{document}
\maketitle
\begin{abstract}
Tone is a prosodic feature used to distinguish words in many languages, some of which are endangered and scarcely documented.
In this work, we use unsupervised representation learning to identify probable clusters of syllables that share the same phonemic tone.
Our method extracts the pitch for each syllable, then trains a convolutional autoencoder to learn a low-dimensional representation for each contour.
We then apply the mean shift algorithm to cluster tones in high-density regions of the latent space. Furthermore, by feeding the centers of each cluster into the decoder, we produce a prototypical contour that represents each cluster.
We apply this method to spoken multi-syllable words in Mandarin Chinese and Cantonese and evaluate how closely our clusters match the ground truth tone categories.
Finally, we discuss some difficulties with our approach, including contextual tone variation and allophony effects.
\end{abstract}

\section{Introduction}

Tonal languages use pitch to distinguish different words, for example, {\em yi} in Mandarin may mean `one', `to move', `already', or `art', depending on the pitch contour. Of over 6000 languages in the world, it is estimated that as many as 60-70\% are tonal \cite{ethnologue, yip-tone}. A few of these are national languages (e.g., Mandarin Chinese, Vietnamese, and Thai), but many tonal languages have a small number of speakers and are scarcely documented. There is a limited availability of trained linguists to perform language documentation before these languages become extinct, hence the need for better tools to assist linguists in these tasks.

One of the first tasks during the description of an unfamiliar language is determining its phonemic inventory: what are the consonants, vowels, and tones of the language, and which pairs of phonemes are contrastive? Tone presents a unique challenge because unlike consonants and vowels, which can be identified in isolation, tones do not have a fixed pitch, and vary by speaker and situation. Since tone data is subject to interpretation, different linguists may produce different descriptions of the tone system of the same language \cite{yip-tone}.

In this work, we present a model to automatically infer phonemic tone categories of a tonal language. We use an unsupervised learning approach: a convolutional autoencoder learns a low-dimensional representation of each tone using only a set of spoken syllables in the target language. This is followed by mean shift clustering to identify clusters of syllables that probably have the same tone. We apply our method on Mandarin Chinese and Cantonese datasets, for which the ground truth annotation is used for evaluation. Our method does not make any language-specific assumptions, so it may be applied to low-resource languages whose phonemic inventories are not already established.

\subsection{Tone in Mandarin and Cantonese}

\begin{figure}
    \centering
    \includegraphics[width=\linewidth]{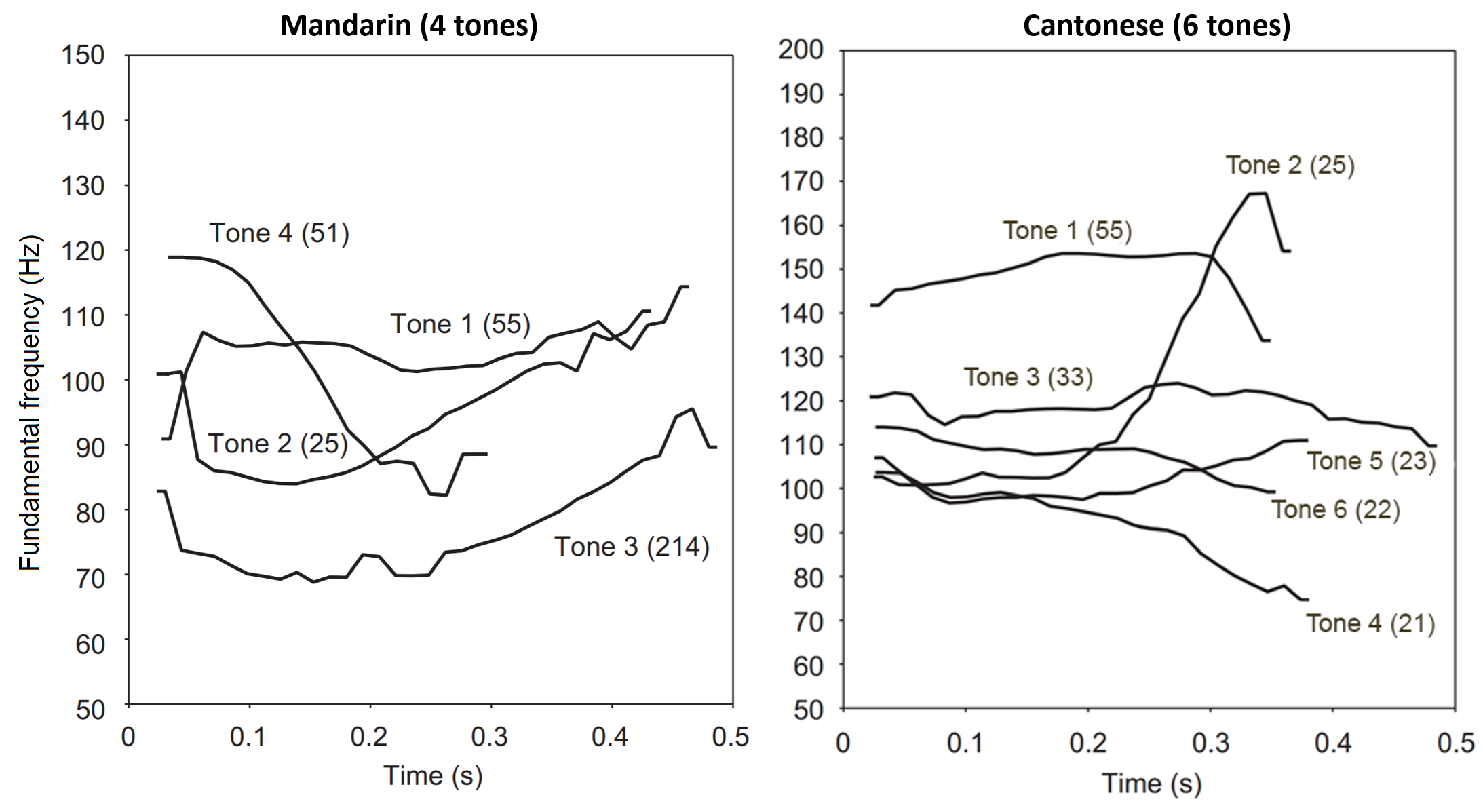}
    \caption{Fundamental frequency (F0) contours for the four Mandarin tones and six Cantonese tones in isolation, produced by native speakers. Figure adapted from \cite{mandarin-cantonese-tones}.}
    \label{fig:mandarin-cantonese-tones}
\end{figure}

\begin{figure*}
    \centering
    \includegraphics[width=0.85\linewidth]{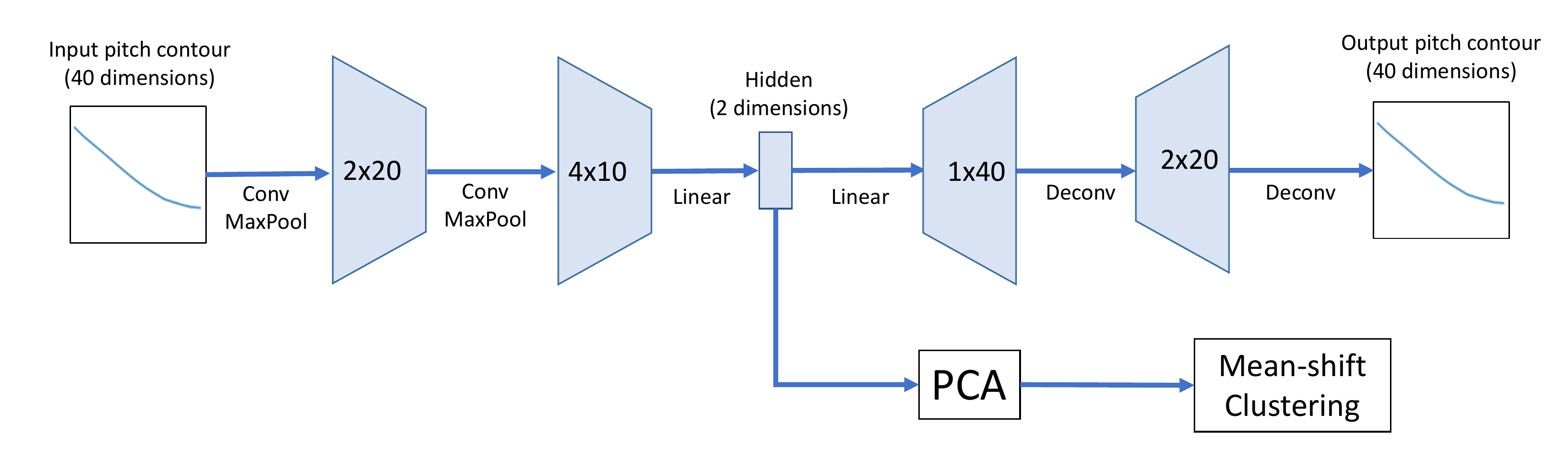}
    \caption{Diagram of our model architecture, consisting of a convolutional autoencoder to learn a latent representation for each pitch contour, and mean shift clustering to identify groups of similar tones.}
    \label{fig:architecture-conv}
\end{figure*}

Mandarin Chinese (1.1 billion speakers) and Cantonese (74 million speakers) are two tonal languages in the Sinitic family \cite{ethnologue}. Mandarin has four lexical tones: high (55), rising (25), low-dipping (214), and falling (51)\footnote{The numbers are Chao tone numerals, where 1 is the lowest and 5 is the highest pitch.}. The third tone sometimes undergoes sandhi, addressed in section \ref{sec:third-tone-sandhi}. We exclude a fifth, neutral tone, which can only occur in word-final positions and has no fixed pitch.

Cantonese has six lexical tones: high-level (55), mid-rising (25), mid-level (33), low-falling (21), low-rising (23), and low-level (22). Some descriptions of Cantonese include nine tones, of which three are {\em checked} tones that are flat, shorter in duration, and only occur on syllables ending in /p/, /t/, or /k/. Since each one of the checked tones are in complementary distribution with an unchecked tone, we adopt the simpler six tone model that treats the checked tones as variants of the high, mid, and low level tones. Contours for the lexical tones in both languages are shown in Figure \ref{fig:mandarin-cantonese-tones}.

\section{Related work}

Many low-resource languages lack sufficient transcribed data for supervised speech processing, thus unsupervised models for speech processing is an emerging area of research. The Zerospeech 2015 and 2017 challenges featured unsupervised learning of contrasting phonemes in English and Xitsonga, evaluated by an ABX phoneme discrimination task \cite{zerospeech-challenge}. One successful approach used denoising and correspondence autoencoders to learn a representation that avoided capturing noise and irrelevant inter-speaker variation \cite{zerospeech-autoencoders}. Deep LSTMs for segmenting and clustering phonemes in speech have also been explored in \cite{muller1} and \cite{muller2}.

In Mandarin Chinese, deep neural networks have been successful for tone classification in isolated syllables \cite{mandarin-classification-cnn} as well as in continuous speech \cite{ryant1, ryant2}. Both of these models found that Mel-frequency cepstral coefficients (MFCCs) outperformed pitch contour features, despite the fact that MFCC features do not contain pitch information. In Cantonese, support vector machines (SVMs) have been applied to classify tones in continuous speech, using pitch contours as input \cite{cantonese-svm}.

Unsupervised learning of tones remains largely unexplored. \citet{levow2006} performed unsupervised and semi-supervised tone clustering in Mandarin, using average pitch and slope as features, and $k$-means and asymmetric $k$-lines for clustering. Graph-based community detection techniques have been applied to group $n$-grams of contiguous contours into clusters in Mandarin \cite{mandarin-tone-shapes}. In recent work concurrent to ours, \citet{fry-thesis} uses adversarial autoencoders and hierarchical clustering to identify tone inventories, and evaluate their method on Mandarin, Cantonese, Fungwa, and English data.

We further explore unsupervised deep neural networks for phonemic tone clustering. It should be noted that our unsupervised model is not given tone labels during training, and the number of tones is assumed to be unknown, so it cannot be directly compared to supervised tone classifiers in the literature.

\section{Data and preprocessing}

We use data from Mandarin Chinese and Cantonese. For each language, the data consists of a list of spoken words, recorded by the same speaker. The Mandarin dataset is from a female speaker and is provided by Shtooka\footnote{\texttt{http://shtooka.net/}, specifically the cmn-caen-tan dataset.}, and the Cantonese dataset is from a male speaker and is downloaded from Forvo\footnote{\texttt{https://forvo.com/}}, an online crowd-sourced pronunciation dictionary. We require all samples within each language to be from the same speaker to avoid the difficulties associated with channel effects and inter-speaker variation. We randomly sample 400 words from each language, which are mostly between 2 and 4 syllables; to reduce the prosody effects with longer utterances, we exclude words longer than 4 syllables.

\label{sec:third-tone-sandhi}

We extract ground-truth tones for evaluation purposes. In Mandarin, the tones are extracted from the pinyin transcription; in Cantonese, we reference the character entries on Wiktionary\footnote{\texttt{https://en.wiktionary.org/}} to retrieve the romanized pronunciation and tones. For Mandarin, we adjust for third-tone sandhi (a phonological rule where a pair of consecutive third-tones is always realized as a second-tone followed by a third-tone), and use the sandhi tone as the ground truth. We also exclude the neutral tone, which has no fixed pitch and is sometimes thought of as a lack of tone.

\subsection{Pitch extraction and syllable segmentation}

We use Praat's autocorrelation-based pitch estimation algorithm to extract the fundamental frequency (F0) contour for each sample, using a minimum frequency of 75Hz and a maximum frequency of 500Hz \cite{praat-f0}. The interface between Python and Praat is handled using Parselmouth \cite{parselmouth}. We normalize the contour to be between 0 and 1, based on the speaker's pitch range.

Next, we manually segment each speech sample into syllables, necessary because syllable boundaries are not provided in our datasets. We sample the pitch at 40 equally spaced points, obtaining a constant length vector as input to our model. Note that by sampling a variable length contour to a constant length, the model does not have information about syllable length; we discuss this design choice in section \ref{sec-limitations}.

\section{Model}

\begin{figure}
\centering
    \includegraphics[width=\linewidth]{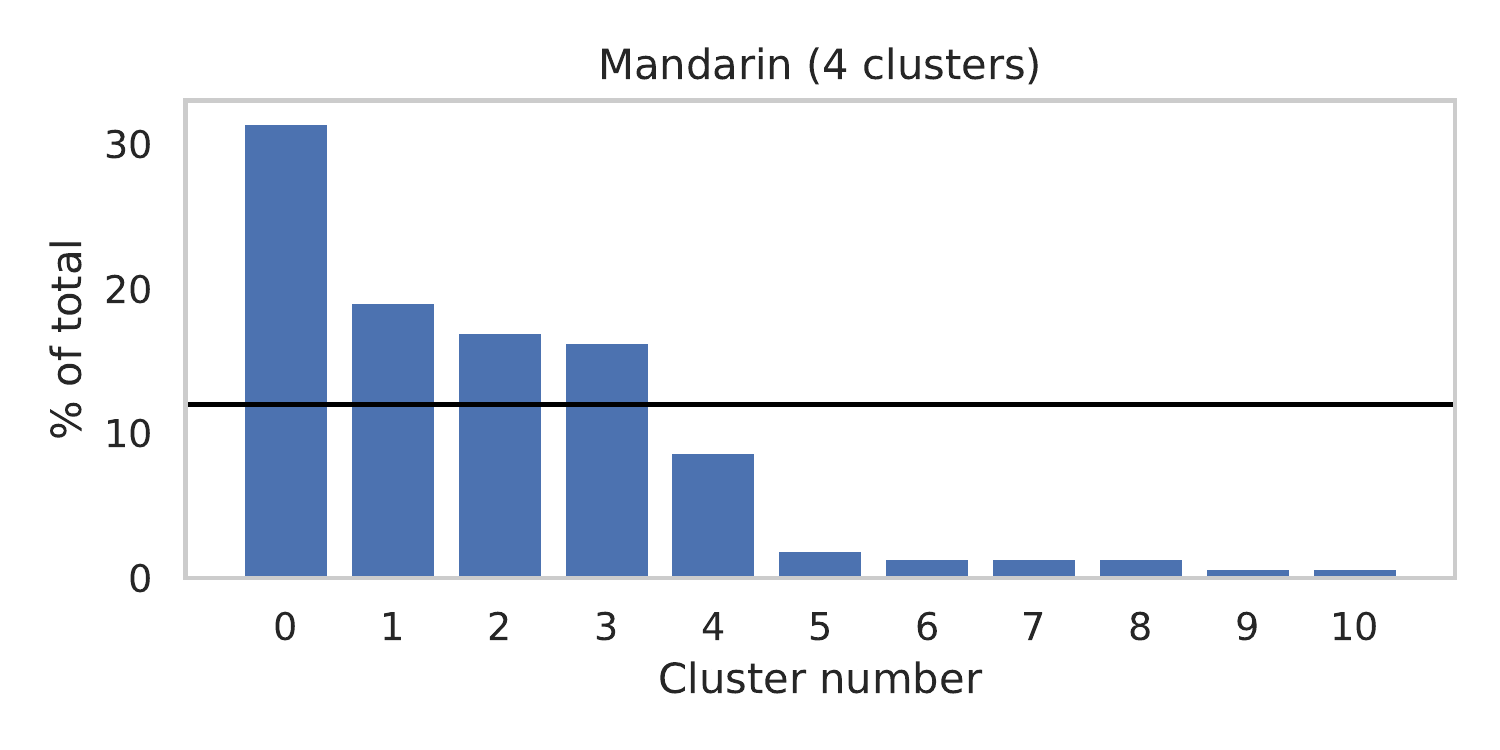}
    \includegraphics[width=\linewidth]{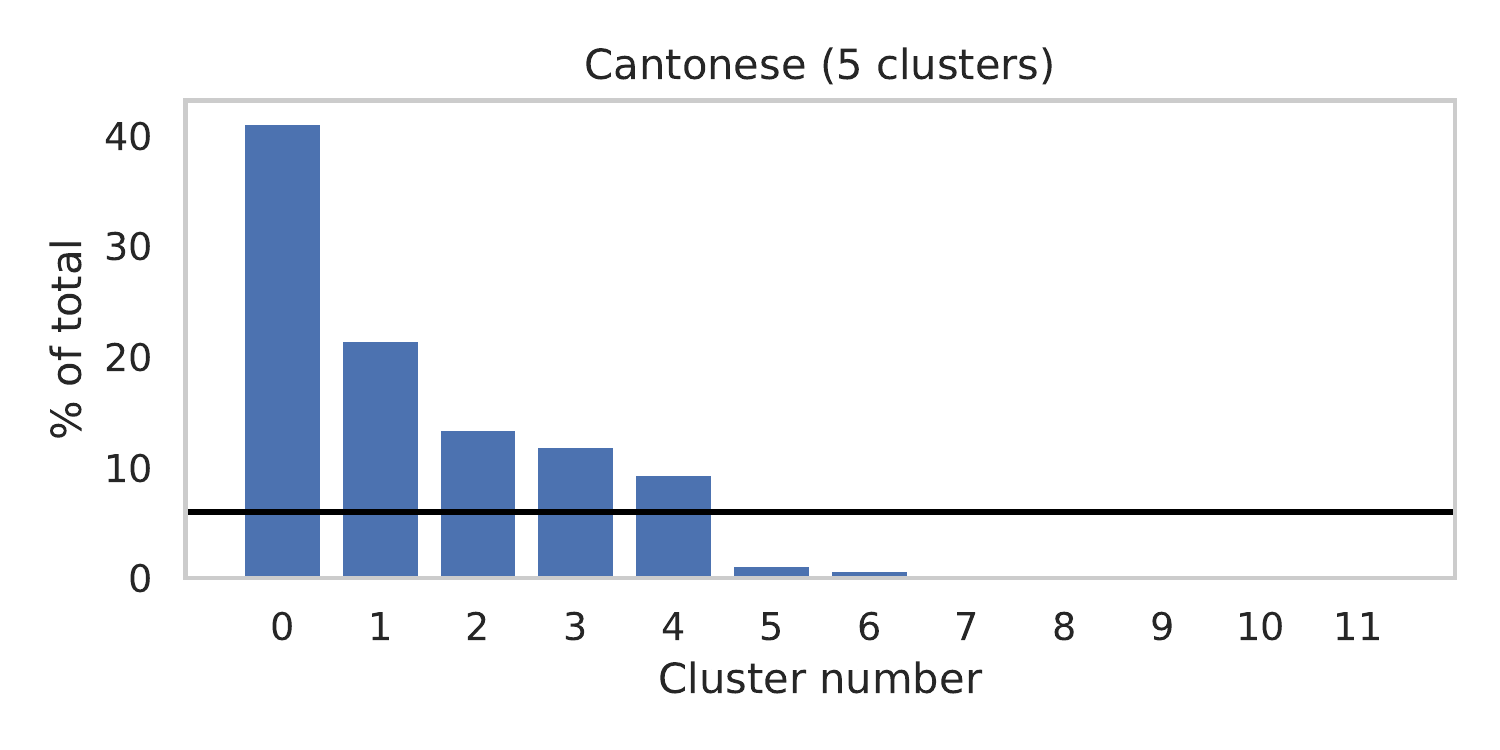}
    \caption{Clusters generated by the mean shift procedure. The black line shows the threshold: we discard clusters with size below this value and treat their points as unclustered.}
    \label{fig:cluster-bars}
\end{figure}

\begin{figure*}
    \centering
    \includegraphics[width=\linewidth]{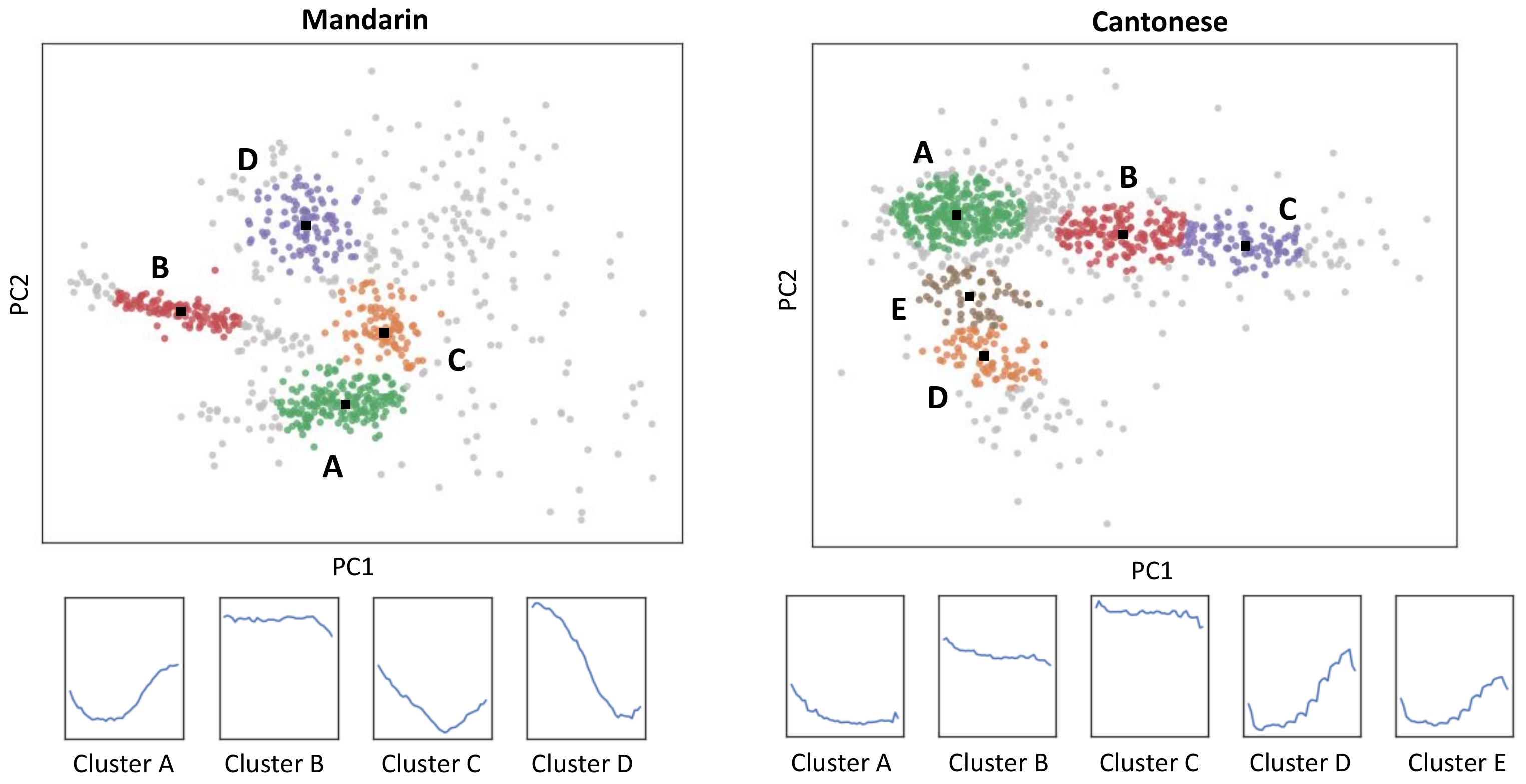}
    \caption{Latent space generated by autoencoder and the results of mean shift clustering for Mandarin and Cantonese. Each cluster center is fed through the decoder to generate the corresponding pitch contour. The clusters within each language are ordered by size, from largest to smallest.}
    \label{fig:representations-contours}
\end{figure*}

\subsection{Convolutional autoencoder}

We use a convolutional autoencoder (Figure \ref{fig:architecture-conv}) to learn a two-dimensional latent vector for each syllable. Convolutional layers are widely used in computer vision and speech processing to learn spatially local features that are invariant of position. We use a low dimensional latent space so that the model learns to generate a representation that only captures the most important aspects of the input contour, and also because clustering algorithms tend to perform poorly in high dimensional spaces.

Our encoder consists of three layers. The first layer applies 2 convolutional filters (kernel size 4, stride 1) followed by max pooling (kernel size 2) and a tanh activation. The second layer applies 4 convolutional filters (kernel size 4, stride 1), again with max pooling (kernel size 2) and a tanh activation. The third layer is a fully connected layer with two dimensional output. Our decoder is the encoder in reverse, consisting of one fully connected layer and two deconvolution layers, with the same layer shapes as the encoder.

We train the autoencoder using PyTorch \cite{pytorch}, for 500 epochs, with a batch size of 60. The model is optimized using Adam \cite{adam} with a learning rate of 5e-4 to minimize the mean squared error between the input and output contours.

\subsection{Mean shift clustering}

We run the encoder on each syllable's pitch contour to get their latent representations; we apply principal component analysis (PCA) to remove any correlation between the two dimensions. Then, we run mean shift clustering \cite{meanshift,ghassabeh}, estimating a probability density function in the latent space. The procedure performs gradient ascent on all the points until they converge to a set of stationary points, which are local maxima of the density function. These stationary points are taken to be cluster centers, and points that converge to the same stationary point belong to the same cluster. We feed the cluster centers into the decoder to generate a prototype pitch contour for each cluster.

Unlike $k$-means clustering, the mean shift procedure does not require the number of clusters to be specified, only a bandwidth parameter (set to 0.6 for our experiments). The cluster centers are always in regions of high density, so they can be viewed as prototypes that represent their respective clusters. Another advantage is that unlike $k$-means, mean shift clustering is robust to outliers.

\subsection{Selecting bandwidth and threshold}

The bandwidth parameter controls the size of the clusters: a higher bandwidth value generates fewer and larger clusters. We tune the bandwidth parameter to produce linguistically plausible tone clusters: we expect between 3 to 8 different clusters, each clusters should have at least 1/10 of the points be assigned to it, and most points should belong to some cluster.

The mean shift procedure assigns every point to some cluster, even if the resulting cluster contains only a few points. Thus, we set a threshold: we treat clusters smaller than the threshold as spurious, and leave their points as unclustered. Figure \ref{fig:cluster-bars} shows the effect of the threshold on both languages.

\subsection{$k$-means baseline}

We implement a simple $k$-means baseline similar to \citet{levow2006}, using two engineered features. The first feature is the average pitch of all the points in the pitch contour; the second feature is the slope of an ordinary least squares regression fit on the pitch contour. After extracting these features for every syllable, we run $k$-means clustering, using the same number of clusters that is chosen by the mean shift algorithm.

\section{Results}

Figure \ref{fig:representations-contours} shows the latent space learned by the autoencoders and the clustering output. Our model found 4 tone clusters in Mandarin, matching the number of phonemic tones (Table \ref{tab:confusion-matrix-mandarin}) and 5 in Cantonese, which is one fewer than the number of phonemic tones (Table \ref{tab:confusion-matrix-cantonese}). In Mandarin, the 4 clusters correspond very well with the the 4 phonemic tone categories, and the generated contours closely match the ground truth in Figure \ref{fig:mandarin-cantonese-tones}. There is some overlap between tones 3 and 4; this is because tone 3 is sometimes realized a low-falling tone without the final rise, a process known as half T3 sandhi \cite{tone-sandhi-book}, thus, it may overlap with tone 4 (falling tone).

In Cantonese, the 5 clusters A-E correspond to low-falling, mid-level, high-level, mid-rising, and low-rising tones. Tone clustering in Cantonese is expected to be more difficult than in Mandarin because of 6 contrastive tones, rather than 4. The model is more effective at clustering the higher tones (1, 2, 3), and less effective at clustering the lower tones (4, 5, 6), particularly tone 4 (low-falling) and tone 6 (low-level). This confirms the difficulties in prior work, which reported worse classification accuracy on the lower-pitched tones because the lower region of the Cantonese tone space is more crowded than the upper region \cite{cantonese-svm}.

\begin{table}[]
    \centering
    \begin{tabular}{|c|c|c|c|c|}
        \hline
        {\bf Cluster} & {\bf T1} & {\bf T2} & {\bf T3} & {\bf T4} \\ \hline
        \bf A & 1 & 163 & 12 & 4 \\ \hline
        \bf B & 108 & 0 & 0 & 1 \\ \hline
        \bf C & 0 & 5 & 53 & 31 \\ \hline
        \bf D & 1 & 0 & 0 & 97 \\ \hline
        \bf N/A & 47 & 30 & 53 & 129 \\ \hline
    \end{tabular}
    \caption{Cluster and tone frequencies for Mandarin.}
    \label{tab:confusion-matrix-mandarin}
\end{table}

\begin{table}[]
    \centering
    \begin{tabular}{|c|c|c|c|c|c|c|}
        \hline
        {\bf Cluster} & {\bf T1} & {\bf T2} & {\bf T3} & {\bf T4} & {\bf T5} & {\bf T6} \\ \hline
        \bf A & 5 & 5 & 59 & 109 & 7 & 105 \\ \hline
        \bf B & 102 & 3 & 36 & 2 & 2 & 7 \\ \hline
        \bf C & 93 & 0 & 0 & 2 & 0 & 0 \\ \hline
        \bf D & 0 & 64 & 4 & 3 & 2 & 11 \\ \hline
        \bf E & 0 & 28 & 2 & 4 & 30 & 2 \\ \hline
        \bf N/A & 70 & 39 & 51 & 45 & 15 & 49 \\ \hline
    \end{tabular}
    \caption{Cluster and tone frequencies for Cantonese.}
    \label{tab:confusion-matrix-cantonese}
\end{table}

\begin{table}[t]
    \centering
    \begin{tabular}{|l|c|c|}
    \hline
     & \multicolumn{1}{l|}{\textbf{Autoencoder}} & \multicolumn{1}{l|}{\textbf{$k$-means}} \\ \hline
     Mandarin (First) & 0.846 & 0.829 \\ \hline
    Mandarin (All) & 0.753 & 0.645 \\ \hline
    Cantonese (First) & 0.575 & 0.493 \\ \hline
    Cantonese (All) & 0.463 & 0.377 \\ \hline
    \end{tabular}
    \caption{Normalized mutual information (NMI) between cluster assignments and ground truth tones, considering only the first syllable of each word, or all syllables.}
    \label{tab:nmi}
\end{table}

To evaluate how much the clusters match the ground truth, we use normalized mutual information (NMI); this is preferable over accuracy because it does not require the number of detected clusters to be the same as the number of tones. In Table \ref{tab:nmi}, we evaluate NMI for our autoencoder model and the $k$-means baseline. We consider two scenarios for each language: using all the syllables (All) and using only the first syllable of each word (First).

In all cases, the clusters from the autoencoder model have higher NMI than the $k$-means model. The improvement is due to the mean shift procedure identifying points that belong to a cluster with high confidence: it only only makes predictions for those points, whereas $k$-means assigns every point to a cluster. All models perform better on the first syllable of each utterance than the rest of the syllables; we discuss the reasons for this in the next section.

\section{Limitations}

\subsection{Contextual effects}

One limitation of our model is it considers syllables in isolation, but in reality, pitch is affected by context. Two types of contextual effects are carry-over and declination. A carry-over effect is when the pitch contour of a tone undergoes contextual variation depending on the preceding tone; strong carry-over effects have been observed in Mandarin \cite{carry-over}. Prior work \cite{levow2006} avoided carry-over effects by using only the second half of every syllable, but we do not consider language-specific heuristics in our model.

Declination is a phenomenon in which the pitch declines over an utterance \cite{yip-tone, cantonese-svm}. This is especially a problem in Cantonese, which has tones that differ only on pitch level and not contour: for example, a mid-level tone near the end of a phrase may have the same absolute pitch as a low-level tone at the start of a phrase.

Contextual effects are apparent in our results (Table \ref{tab:nmi}). In both Mandarin and Cantonese, the clustering is more accurate when using only the first syllable (which is not affected by carry-over or declination), compared to using all the syllables.

\subsection{Minimal pairs and allotones}

\label{sec-limitations}

Tone is not a purely phonetic property: it is impossible to determine, from phonetics alone, whether two pitch contours have the same or different tones. The same underlying tone may manifest as several different allotones depending on the phonetic context.

An example of this appears in Cantonese. Its tone system is sometimes analyzed as having nine tones instead of six, where six of the tones are only permitted in open syllables (e.g. {\em si}) and three are only permitted in checked syllables (e.g. {\em sik}). Other analyses use a six-tone system, treating the three checked tones as allotonic variants of the high, mid, and low tones. By taking this approach, one implies that length is a property of the syllable and cannot be solely responsible for contrasting two tones.

Length is not the only differentiating factor for allotones. Another example is in Wu Chinese, where syllables beginning with voiced consonants have lower pitch than those beginning with voiceless consonants \citep{yip-tone}. Thus the same language may have vastly different numbers of tones, depending on the analysis.

Linguistically, two phonemic tones are considered to be contrastive if there exists a minimal pair: two semantically different lexical items that are identical in every aspect except for tone. This definition is the most widely used because it clearly settles disagreements about whether two tones are same or different. However, it is problematic for unsupervised models that only have access to phonetic and not semantic information. This issue is not unique to tone: similar difficulties have been noted when attempting to identify consonant and vowel phonemes automatically \citep{phoneme-is-hard}.

\section{Conclusion}

We propose a model for unsupervised clustering and discovery of phonemic tones in tonal languages, using spoken words as input. Our model extracts the F0 pitch contour, trains a convolutional autoencoder to learn a low-dimensional representation for each contour, and applies mean shift clustering to the resulting latent space. We obtain promising results with both Mandarin Chinese and Cantonese, using only 400 spoken words from each language. Cantonese presents more difficulties because of its larger number of tones, especially at the lower half of the pitch range, and also due to multiple contrastive level tones. Still, in both our languages, our method finds clusters of tones that better match the ground truth than the $k$-means baseline. Finally, we discuss the effects of contextual variation and the limitations of unsupervised learning for the tone induction problem.

\section{Acknowledgments}
We thank Prof Gerald Penn for his help suggestions during this project. Rudzicz is a CIFAR Chair in AI.

\bibliography{anthology,acl2020}
\bibliographystyle{acl_natbib}

\end{document}